# Multiple Access for Visible Light Communications: Research Challenges and Future Trends

Sarah S. Bawazir, Student Member, IEEE,

Paschalis C. Sofotasios, Member, IEEE, Sami Muhaidat, Senior Member, IEEE,

Yousof Al-Hammadi and George K. Karagiannidis, Fellow, IEEE

Abstract

The ever-increasing demand of mobile Internet and multimedia services poses unique and significant challenges for current and future generation wireless networks. These challenges are mainly relating to the support of massive ubiquitous connectivity, low latency and highly efficient utilization of spectrum resources. Therefore, it is vital to consider them extensively prior to design and deployment of future wireless networks. To this end, the present article provides a comprehensive overview of a particularly promising and effective wireless technology, namely, visible light communication (VLC). In this context, we initially provide a thorough overview of frequency domain multiple access techniques for single and multi-carrier systems, which is then followed by an in depth discussion on the technical considerations of optical code division multiple access techniques and their adoption in indoor VLC applications. Furthermore, we address space division multiple access and, finally, we revisit and analyze the distinct characteristics of a new technology, namely, non-orthogonal multiple access, which has been proposed recently as a particularly effective solution.

## I. INTRODUCTION

Visible light communication (VLC) is an evolving communication technology that has been proposed as a promising candidate for high speed communications, particularly in indoor environments. In VLC, the lighting infrastructure can be used to provide both illumination and connectivity. This is due to the significant advancement in light emitting diodes (LEDs) technology [1], which enabled the variation of the intensity of light at very high frequencies making data transmission over light possible without being perceived by the human eye. This process is known as intensity modulation (IM). At the receiving end and in a process called direct detection (DD), a photodetector or an image sensor is used to detect the changes of the light intensity [1].

*A. Features and Limitations of VLC*

VLC is an attractive alternative to conventional radio frequency (RF) communication for indoor environments [2], particularly in demanding scenarios with increased quality of service requirements. Besides providing high speed connectivity and unlimited bandwidth, VLC technology has the following distinct advantages:

(a) VLC spans 389 terahertz of unregulated spectrum, i.e. from 400 THz (red) up to 789 THz (violet), which lies in the visible light range of the electromagnetic spectrum.

(b) Wavelengths corresponding to visible light frequencies have been shown to be safe to the human body.

(c) IM/DD-based VLC systems consist of relatively inexpensive optoelectronic devices at both transmitter and receiver sites, namely, LEDs and photodiodes.

(d) Visible light cannot penetrate through walls and objects which enables the following:
- The use of LEDs as small cells that can provide high quality services without inter-cell interference.
- Inherently secure wireless communication, i.e. no eavesdropping.

Nevertheless, despite the above remarkable advantages, one of the major challenges in VLC is the limited modulation bandwidth of the light sources [3]. This is technically a hardware constraint that constitutes the main challenge in the development of high data rate VLC links. However, this issue can be effectively resolved with the aid of multiple access techniques, high order modulation schemes and frequency reuse.

*B. Conventional and Emerging Multiple Access Schemes*

Radio communication-based multiple access (MA) techniques have demonstrated a distinct potential in efficiently sharing the available network resources among a large number of users. This has been typically realized by means of frequency or time division, which constituted frequency division multiple access (FDMA) and time division multiple access (TDMA) schemes, which were employed in the first (1G) and second (2G) generations of mobile communications, respectively. On the contrary, 3G was based on code division multiple access (CDMA), which allowed the assignment of frequency bands and time intervals to different users with a distinct code that provided immunity from the resulting interference. In the context of power domain, multiple users can share time and frequency resources simultaneously by means of power domain multiplexing. A typical example is non-orthogonal multiple access (NOMA), which is a power domain multiplexing technique where multiple users send and receive over the same frequency band at the same time interval using varying power levels. It has been shown that this technique has the potential to increase system throughput and support ubiquitous connectivity.

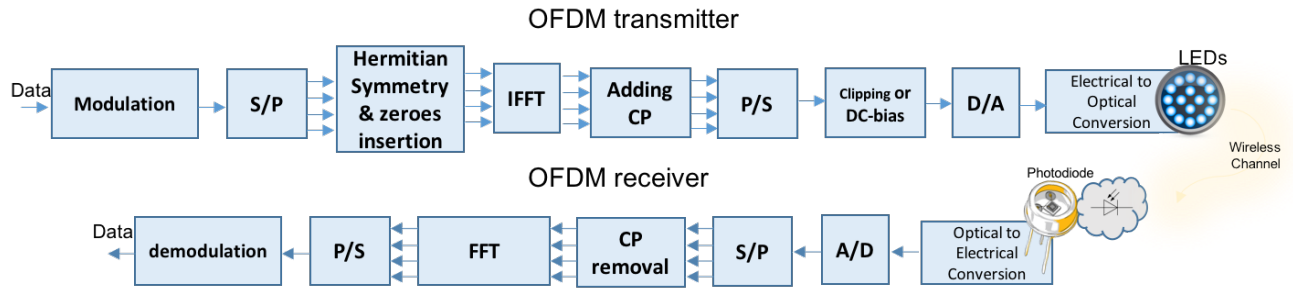

Fig. 1: Optical OFDM transmitter and receiver.

It is noted that the majority of conventional MA schemes proposed for RF systems cannot be applied to VLC without essential modifications. For example, IM and DD systems require real and non-negative information signals [3], hence, intermediate processing should be considered at the transmitter in order to transform complex symbols into real and positive valued symbols. A summary of these processes is presented in the following section. Based on the above, the aim of this paper is to provide a comprehensive review of multiple access techniques that are capable of facilitating increased data rates and bandwidth efficiency in the context of VLC systems. This review also includes a thorough presentation of the current technical challenges and constraints relating to these techniques providing essential insights that are expected to be useful in the design and deployment of such systems.

## II. Optical Frequency Domain Multiple Access Techniques

Orthogonal frequency division multiplexing (OFDM) was introduced as a key enabler for high speed communications. In the context of VLC systems, as data rate increases the limited bandwidth of the involved LEDs leads to inter-symbol interference (ISI), which can be minimized with the aid of effective equalization-enabled techniques. Also, OFDM is an efficient technique in terms of ISI as it is based on the use of a large number of low rate orthogonal subcarriers to achieve increased throughputs [2], [4]. It is recalled that OFDM-based receivers require a simple one tap equalizer [5] which reduces the complexity of the system. Fig. 1 illustrates the transceiver structure of a typical OFDM system. However, there are certain limitations in OFDM realization in the context of VLC systems that cannot be considered negligible. Specifically, conventional OFDM generates complex and bipolar waveforms which are not compatible with the IM/DD schemes employed in VLC. This issue can be effectively resolved by exploiting the Hermitian symmetry on the subcarriers in the frequency domain to produce a real OFDM signal, which is suitable for VLC systems. However, the resulting signal is still bipolar; as a result, several new schemes that provide unipolar and real time-domain OFDM signals were proposed for ensuring compatibility with VLC systems. In more details, the following techniques have been reported for resolving the aforementioned issue of the involved bi-polarity:

1) DC-biased Optical OFDM.
2) Asymmetrically Clipped Optical OFDM.
3) Asymmetrically Clipped DC-biased Optical OFDM.
4) Polar OFDM.

As already mentioned, MA schemes constituted a core part of all generations of telecommunication systems as FDMA, TDMA and CDMA were adopted in 1G, 2G and 3G, respectively. Likewise, orthogonal frequency division multiple access (OFDMA) is an MA scheme that is based on OFDM and was adopted in 4G systems. Based on its distinct advantages, OFDMA has been also considered as an effective technology in optical communication systems. The corresponding OFDM-based MA schemes used in the context of VLC are as follows:

1) Optical Orthogonal frequency-division multiple access (OFDMA).
2) Optical OFDM-interleave division multiple access (OFDM-IDMA).

It is noted here that OFDMA is considered, naturally, an extension of OFDM where each user is allocated a group of subcarriers in each time slot, depending on the corresponding traffic requirements and channel conditions. Furthermore, it is noted that OFDM-interleave division multiple access (OFDM-IDMA) is a hybrid scheme between OFDM and IDMA technologies.

*A. OFDMA versus OFDM-IDMA*

In what follows, we provide a comparison between OFDMA and OFDM-IDMA schemes. To this end, it is recalled that both multi-carrier schemes are asymmetrically clipped at zero, after the OFDM modulation, in order to become compatible with IM/DD. Users in IDMA schemes are separated by distinct random chip-level interleaving. Based on this, an iterative chip by chip multiuser detection is employed at the receiver site in order to recover the data of each user. In this context, it was shown in [6] that OFDM-IDMA outperforms OFDMA when the signal-to-noise ratio (SNR) is above the 10 dB threshold for a system with modulation size of M = 16. It is also shown that optical OFDM-IDMA exhibits the advantage of being more power efficient, compared to optical OFDMA, while its decoding complexity is linear with respect to the number of users and independent from the number of paths [6]. On the contrary, optical OFDMA exhibits the advantage of a lower decoding complexity and a slightly lower peak-to-average power ratio (PAPR) compared to optical OFDM-IDMA schemes.

*B. Single-Carrier FDMA*

Single-carrier frequency division multiple access (SC-FDMA) is a multiple access scheme that is based on the concept of frequency division and its main advantage is the relatively low PAPR. Its distinct usefulness is also evident by the fact that its adoption in the context of VLC can assist in effectively overcoming the LED nonlinearities. To this end, Kashef et al. [7] investigate the performance of SC-FDMA for a multiuser data transmission in a VLC setting consisting of a VLC access point and multiple VLC receiving devices. Based on this, they optimize the sub-carrier power allocation and the total transmission power, in order to maximize the minimum achievable data rate under the transmission power constraints imposed by room lighting requirements. The access point considers SC-FDMA to overcome the clipping effect that results from the limited dynamic range of the LED. Furthermore, each receiving terminal is equipped with minimum mean square error (MMSE)-based linear frequency domain equalizer (FDE). It is also recalled here that high PAPR and the limited dynamic range of the components of the system constitute the main challenges affecting OFDM-based schemes in VLC systems.

III. OPTICAL CODE DOMAIN MULTIPLE ACCESS TECHNIQUES

Code division multiple access (CDMA) is a technique which assigns a dedicated code to each user to enable simultaneous transmission or reception over the same frequency band. At the receiving end, the user correlates the received signals with their designated code and proceeds in the decision accordingly. Optical CDMA (O-CDMA) operates in a similar manner as RF-based CDMA with the difference that O-CDMA uses optical codes proposed by [8] and termed as *optical orthogonal codes (OOCs)*. It is emphasized in [8] that the performance of optical CDMA is bounded by the choice of the high rate signature sequences. Also, in order to guarantee optimal recovery at the receiver, the codes must satisfy the following two conditions:

1) Each code can be clearly distinguished from a shifted version of itself.
2) Each code should be clearly distinguished from a possibly shifted version of every other sequence in the set.

As stated in [9], these codes are harder to generate for a large number of users. Motivated by this, Guerra-Medina et al. [9] proposes random optical codes (ROCs) in order to spread the bandwidth of each signal. However, although ROCs are relatively easier to generate compared to OOCs, they are practically suboptimal as their simplicity comes at the expense of performance degradation.

The analysis in [10] considers color-shift-keying alongside CDMA to provide a high capacity multiple access system. Importantly, it is proved that the proposed system exhibits a 3-dB transmission gain for each user compared to on-off keying (OOK). In the same context, the contribution in [11] discusses a multi-carrier (MC) CDMA-based indoor optical wireless scheme that can be considered a hybrid between OFDM and CDMA techniques. In MC-CDMA, the data symbols in the frequency domain of each user are spread over OFDM subcarriers and the aggregate sum of the frequency domain symbols from the different users is subsequently represented in the time domain through OFDM modulation. In addition, it was shown in [11] that sub-carrier selection can be utilized to reduce the average-transmit optical power in MC-CDMA along with sub-carrier selection criteria for both pre- and post-equalization. However, there is also a number of limitations in the O-CDMA scheme. For example, in order to achieve optimal performance, longer OOC signature sequences are generated at the expense of reducing the achievable data rates [1]. Based on this, a spectrally efficient scheme was reported in [1], and references therein, where M-ary information is transmitted by utilizing different cyclic shifts of the assigned sequence. This is known as code cycle modulation (CCM). However, besides the usefulness of this approach, its main challenge is its inability to support dimming and thus, it requires the design of complex receivers [1].

## IV. Optical Spatial Division Multiple Access Techniques

In a space division multiple access (SDMA) system, spatial diversity is employed to provide shared frequency-time resources among a group of users. In a conventional RF-SDMA system, an antenna array is used to simultaneously generate a number of narrow beams corresponding to the locations of active users, which enables multiple users to be served at the same time slot [12]. However, it is recalled that SDMA-based transmission in RF requires multiple chains and complex beam-steering algorithms [12] that are not necessary in VLC systems. The underlying reason is that LEDs have limited field-of-views (FOVs) inherently, which enables them to generate directional light beams as illustrated in Fig. 2.

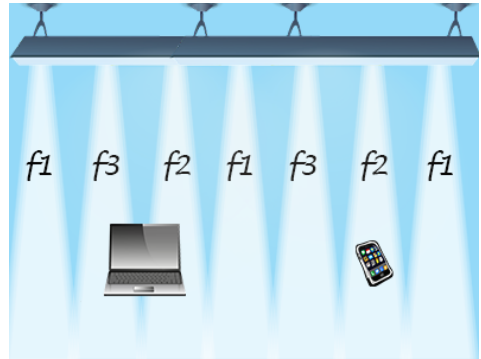

Fig. 2: Space Division Multiple Access.

The analysis in [12] utilizes an angle diversity transmitter to enable parallel transmissions. It was shown that when the number of transmitting elements increases, the improvement of the throughput increases significantly. In addition, it was demonstrated that an optical SDMA system can improve the system throughput over ten times more than an optical TDMA system [12]. In the same context, Yin et al. [13] proposed a low-complexity suboptimal algorithm for a coordinated multi-point (CoMP) VLC system with SDMA grouping. It was shown that this algorithm, i.e. random pairing (RP), improves the system performance and fairness-throughput trade-off. The comparison was conducted against two distinct baselines, namely, a system employing coordination with no SDMA grouping and a system employing FDMA with no coordination. To illustrate, Fig. 3a demonstrates that the fairness throughput, which is measured by Jain's index of fairness (JIF), of the proposed RP algorithm is higher than the one corresponding to the CoMP system without SDMA grouping, for different user loadings, i.e. number of user equipment (UEs) in an attocell. On the contrary, the fairness-throughput of FDMA with no frequency reuse is higher than that of the other two systems. Fig. 3b illustrates the performance of the three systems in terms of the area spectral efficiency (ASE). It is observed that the performance of CoMP without SDMA grouping is high for low user loadings, yet, as the number of user loadings increase, the system exhibits a significant reduction of the ASE. On the contrary, a system employing the RP algorithm is more robust to the increase in the number of UE's and maintains higher performance. Finally, it is evident that the FDMA VLC system has low efficiency compared to other systems due to the lack of frequency reuse [13].

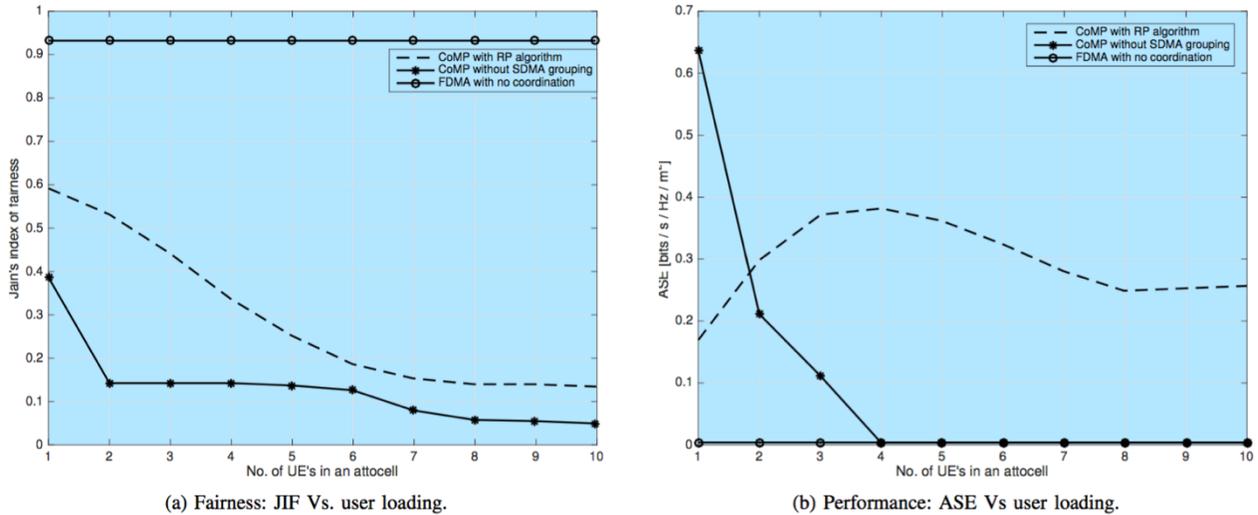

Fig. 3: Performance and fairness trade-off in CoMP with and without SDMA grouping [13].

## V. Optical Non-orthogonal Multiple Access

In this section, we focus on non-orthogonal multiple access (NOMA), which is expected to increase system throughput and accommodate ubiquitous connectivity in VLC systems. NOMA enables users to access communication channels simultaneously by allowing each of them to access the entire bandwidth simultaneously. This can be realized through employing the principle of power-domain multiplexing (PDM), in which different users are allocated different power levels, depending on the corresponding channel conditions [14].

It is recalled that NOMA was proposed for future radio access [14] in order to achieve high capacity gains and system throughput. Since its inception, this concept was intended for RF systems; however, it can also be applicable and arguably more valuable in VLC systems for the following reasons:

1) It is flexible and efficient in multiplexing a small number of users, which could be a limitation when it comes to RF system requirements. However, this does not constitute an issue in indoor VLC systems because a single light fixture can only serve users located in its close proximity, which practically limits the number of users requiring service.

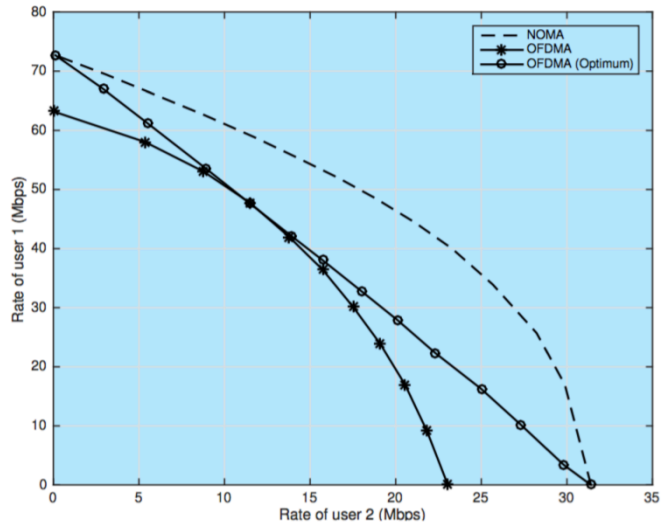

Fig. 4: Boundary of rate pairs comparison between NOMA and OFDMA (assumption: perfect interference cancellation) [15].

2) NOMA receivers rely on channel state information (CSI), which can be estimated relatively accurately in a VLC system due to the fundamentally low mobility of indoor wireless devices.

## A. Operation of NOMA

The basic principle underlying NOMA is the superposition of signals in the power domain. To this end, more power is allocated to the farthest users in order to compensate for the corresponding degradations due to the distant location from the transmitter [14]. To this effect, NOMA-based receivers perform successive interference cancellation (SIC), which practically exhibits a relatively low decoding complexity [14]. This process is essential for enabling each user to extract its own information signals without interference from signals intended for other users. Yet, effective SIC requires accurate knowledge of CSI in order to determine the allocated power for each user, and the decoding order [3]. Such information is readily accessible in indoor VLC due to the practically deterministic nature of the channel. In fact, this is the core advantage of NOMA in the context of VLC over RF-based NOMA, which is also evident by the fact that SIC performs better with a small number of users [15].

*B. NOMA versus OFDMA*

Given the distinct advantages of NOMA and OFDMA, it is considered essential to provide a comparison between these two technologies leading to useful insights on their applicability in the context of VLC systems. Based on this, NOMA was thoroughly analyzed in [15] and compared against OFDMA in the context of a VLC downlink scenario with two users. DC-biased Optical OFDM (DCO-OFDM) was adopted for both NOMA and OFDMA to convert the bipolar signal into a unipolar signal. It was demonstrated that when perfect interference cancellation is assumed, NOMA achieves higher data rates for both users except for the case that the rates are equal to the single user capacity, as illustrated in Fig. 4. On the contrary, NOMA's performance begins to deteriorate when interference cancellation becomes imperfect. For example, for a cancellation error of 1%, NOMA still outperforms its OFDMA counterpart; nevertheless, when the involved error is greater than 2%, NOMA's performance degrades rapidly. In the same context, Marshoud et al. [3] developed a framework for a downlink NOMA scheme in an indoor multi-LED VLC system. This work utilizes a novel channel-dependent gain ratio power allocation (GRPA) strategy in order to maximize the corresponding users' sum rate. Unlike the static power allocation approach, GRPA considers the channel conditions for each user in order to allocate power fairly and efficiently [3]. Based on this, the amount of power allocated to each user depends not only on the channel gain of that user but also on its decoding order as well as the gain of the first sorted user [3]. To this effect, it is shown that GRPA compensates for the channel difference among users, leading to a higher capacity for the same channel conditions as compared with the static power allocation method. For example, at a BER of $10^{-3}$, using the static power allocation results in accommodating 4 users, whereas using GRPA results in accommodating 6 users.

The performance of NOMA can be enhanced further in VLC by tuning the receiver's field of view (FOV) and the transmission angles [3], [16]. It is shown in [3] that for a small number of users, i.e. less than five users, tuning the transmission angles and FOV's results in increasing the sum data rate. On the contrary, as the number of users increases, the power allocated to each user decreases. Therefore, it is evident that tuning the transmission angles puts the users at the edge of the cell at a disadvantage, i.e. the cell edge users in this case can only receive from one transmitter. As a consequence, transmission angles tuning in this case degrades the overall system throughput. Finally, when comparing TDMA with NOMA, it was shown in [16] that the latter is capable of increasing the system capacity by 125%, if LEDs with 30° semi-angle ($\varphi_{1/2}$) are used. This is illustrated in Fig. 5, while it is noted that the larger the LEDs' semi-angle, the higher the coverage probability.

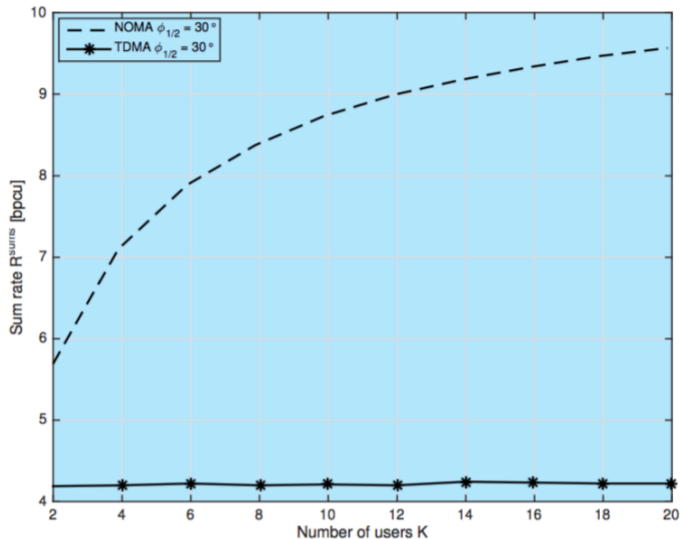

Fig. 5: Ergodic sum rate for different number of users [16].

*C. MIMO NOMA in VLC*

As mentioned earlier, tuning the FOVs and transmission angles in VLC systems allows for two additional degrees of freedom that ultimately improve the overall system performance. Therefore, applying MIMO to NOMA brings additional degrees of freedom to improve the system performance [17]. This was analyzed extensively in radio communications, e.g. [17] and the references therein. There, it was shown that combining MIMO and NOMA can lead to high spectral efficiency and throughput. Based on this, different scenarios were analyzed in the RF literature. Specifically, the following scenarios were investigated in [17] in the context of RF-based MIMO NOMA systems:

1) A base station with multiple antennas and users with a single antenna.
2) A multiple-antenna base station and two multiple-antenna users.
3) A clustering based scenario where users are grouped into small-size groups with NOMA implemented in each group and MIMO detection used to cancel inter-group interference, i.e. inter-cluster interference.

MIMO NOMA can be combined with conventional orthogonal multiple access (OMA) techniques by dividing the system into multiple groups and using NOMA within each group. The contribution of [17] presents a general (uplink/downlink) framework for MIMO-NOMA in RF-based system with higher diversity. Likewise, Choi et al. [18] investigate optimal power allocation for a layered transmission of MIMO-NOMA and show that the main advantage of a layered transmission lies in the linearly growing complexity of detection/decoding at the users' end with the growing number of transmit antennas (or layers).

To the best of our knowledge, no implementations in the context of VLC systems have been reported in the open technical literature. Therefore, considering the above scenarios in VLC is important in obtaining higher data rates, more capacity and higher spectral efficiency. As a result, it is essential to investigate the corresponding theoretical and technical challenges, and constraints prior to design and deployment of future optical wireless communication systems.

*D. Open Problems and Potential Solutions*

It is evident that the required SIC process in NOMA based systems imposes intensive computational requirements as each receiver has to perform SIC in order to decode the signal that is intended for it. Thus, the corresponding receiver complexity can be a drawback to NOMA [15], while its performance is susceptible to small SIC errors. Thus, in depth investigations should be carried out on how to address the efficiency and sensitivity of SIC receivers. Furthermore, more practical investigations on improving multiuser power allocation algorithms are necessary along with extensive tests on the corresponding performance in the case of large numbers of users. Also, the relationship between FOV and both the decoding order and the distance of the user from the transmitter, should be characterized and quantified. Finally, realistic mobility scenarios in VLC systems should be analyzed in detail to ensure accurate and reliable results that will be useful in the successful deployment and operation of VLC systems.

## VI. CONCLUSION

In this paper, the major multiple access techniques for visible light communication systems were reviewed, namely, O-FDMA, O-OFDM-IDMA, SC-FDMA, O-CDMA, SDMA and NOMA, alongside discussions and suggestions on the extension of an uprising RF scheme, MIMO- NOMA, in the context of VLC systems. The main advantages and disadvantages of optimal multiple access techniques have been revisited along with the selection criteria for each considered scenario. In general, the main challenges or limitations that are reported in this paper are related to transmitter, receiver and decoder complexity, spectral efficiency and power allocation which are considered particularly useful in the effective design and deployment of efficient and robust VLC systems.